\documentclass[reprint,
amsmath,amssymb,aps,showkeys,showpacs,
twoside,final,secnumarabic,
nofootinbib]{revtex4-2}

\usepackage[paperwidth=205mm,paperheight=290mm,top=17mm,bottom=25mm,
inner=17mm,outer=17mm,
twoside]{geometry}

\usepackage{cmap} 
\usepackage[T1,T2A]{fontenc}
\usepackage[utf8]{inputenc}
\usepackage[russian,english]{babel}
\usepackage{color}
\usepackage{graphicx}
\usepackage{dcolumn}
\usepackage{bm} 
\usepackage[unicode=true,colorlinks=true,linkcolor=magenta, urlcolor=blue, citecolor = blue,breaklinks]{hyperref}
\usepackage{multirow}
\usepackage{url}
\usepackage{breakurl}
\DeclareGraphicsExtensions{.eps}
\usepackage{mathrsfs}

\newcommand{\be}{\begin{equation}}
\newcommand{\ee}{\end{equation}}
\newcommand{\ba} {\begin{equation}\begin{aligned}}
\newcommand{\ea} {\end{aligned}\end{equation}}

\newcommand{\sL}{\mathscr{L}}

\newcommand{\cM}{\mathcal{M}}

\newcommand{\hc}{\text{h.c.}}
\newcommand{\nn}{\nonumber}

\newcommand{\ov}[1]{\overline{#1}}

\def\tH{\widetilde{H}}

\newcommand{\keV}{\ \text{keV}}

\newcommand{\GeV}{\ \text{GeV}}

\newcommand{\PQ}{U(1)_\text{PQ}}

\begin{document}

\title{Mass Generation for Axion Like Particles}
\thanks{Proceedings of the 22nd Lomonosov Conference on Elementary
Particle Physics (Moscow, Russia, August 21--27, 2025)}

\author{\firstname{Luca}~\surname{Merlo}}
\email[E-mail: ]{luca.merlo@uam.es}
\affiliation{Departamento de F\'isica Te\'orica and Instituto de F\'isica Te\'orica UAM/CSIC,\\
Universidad Aut\'onoma de Madrid, Cantoblanco, 28049, Madrid, Spain}

\begin{abstract}
The most widely accepted solution to the Strong CP problem is the QCD axion, whose traditional parameter space is relatively constrained. However, recent studies on exotic axion models have shown that this space can be significantly extended, opening up a wide range of previously unexplored possibilities for model building. I first review the current status of axions and axion-like particles, then focus on the challenges related to the origin of their masses and explore potential connections to other unresolved issues within the Standard Model.
\end{abstract}


\maketitle


\section{Introduction}\label{intro}

Axions, together with the more general class of axion-like particles (ALPs), have been undergoing a revival of interest in recent years.  The axion solution to the Strong CP problem relies on the possibility of  redefining away the so-called QCD-$\theta$ parameter. In the original formulations of the late 1970s~\cite{Peccei:1977hh,Weinberg:1977ma,Wilczek:1977pj}, this mechanism is associated with a global Abelian symmetry, the Peccei--Quinn (PQ) symmetry $\PQ$. The spontaneous breaking of $\PQ$ gives rise to a Goldstone boson (GB), the axion $a$. 

For this solution to be effective, the PQ symmetry must be anomalous with QCD. As a consequence, the axion acquires a non-perturbative mass $m_a$, which is inversely proportional to its decay constant $f_a$, $m_a\, f_a \sim 0.01 \, \GeV^2$.

For almost four decades, QCD axion models~\cite{Peccei:1977hh,Weinberg:1977ma,Wilczek:1977pj,
Zhitnitsky:1980tq,Dine:1981rt,Kim:1979if,Shifman:1979if} were confined to a very narrow region in the $(m_a, f_a)$ parameter space. However, more recent work (See Ref.~\cite{deGiorgi:2024elx} and references therein) has shown that the relation between $m_a$ and $f_a$ can be relaxed, while still solving the Strong CP problem. 

This represents a turning point in axion physics, offering strong motivation for collider searches of 
not-so-light pseudo-scalars, commonly referred to as ALPs.

The definition of an ALP in the literature is not unique. In this work, we consider an ALP as a pseudo-scalar with (dominant) derivative couplings, whose mass is not necessarily associated with any non-perturbative QCD effect. According to the new theoretical results mentioned above, an ALP may or may not be related to a solution of the Strong CP problem. 

Numerous realizations of ALPs have been proposed in the literature, including: 
ALPs associated with flavour dynamics~\cite{Davidson:1981zd,Wilczek:1982rv,Ema:2016ops,Calibbi:2016hwq,Arias-Aragon:2017eww,Arias-Aragon:2022ats,DiLuzio:2023ndz};
ALPs connected to neutrino mass generation~\cite{Chikashige:1980qk,Chikashige:1980ui,Gelmini:1980re};
ALPs appearing in composite Higgs models~\cite{Merlo:2017sun,Brivio:2017sdm,Alonso-Gonzalez:2018vpc,Alonso-Gonzalez:2020wst};
ALPs in supersymmetric contexts~\cite{Bellazzini:2017neg};
ALPs arising in string theory frameworks~\cite{Witten:1984dg,Choi:2006qj,Svrcek:2006yi,Arvanitaki:2009fg,Cicoli:2012sz};
ALPs as Dark Matter candidates~\cite{Gelmini:1984pe,Berezinsky:1993fm,Lattanzi:2007ux,Bazzocchi:2008fh,Lattanzi:2013uza,Queiroz:2014yna};
ALPs playing a role in cosmological observables~\cite{Ferreira:2018vjj,DEramo:2018vss,Escudero:2019gvw,Arias-Aragon:2020qtn,Arias-Aragon:2020qip,Arias-Aragon:2020shv,Ferreira:2020bpb,Escudero:2021rfi,Araki:2021xdk,DEramo:2021psx,DEramo:2021lgb,DEramo:2022nvb}.

Given the wide variety of theoretical contexts in which ALPs appear, part of the community has pursued the development of an Effective Field Theory (EFT) framework to encode their generic features. Following the seminal work of Ref.~\cite{Georgi:1986df}, several studies have defined the ALP effective Lagrangian~\cite{Choi:1986zw,Salvio:2013iaa,Brivio:2017ije}, with the goal of exploring possible signals at low-energy facilities~\cite{Izaguirre:2016dfi,Merlo:2019anv,Aloni:2019ruo,Bauer:2019gfk,Bauer:2020jbp,Bauer:2021mvw,Carmona:2021seb,
Guerrera:2021yss,Gallo:2021ame,Bertholet:2021hjl,Cheng:2021kjg,Bonilla:2022qgm,Bonilla:2022vtn,deGiorgi:2022vup,Guerrera:2022ykl,
Bonilla:2023dtf,Arias-Aragon:2023ehh,DiLuzio:2024jip,deGiorgi:2024str,Alda:2024cxn,Alda:2024xxa,
Arias-Aragon:2024qji,Arias-Aragon:2024gdz,Bisht:2024hbs,MartinCamalich:2025srw,Alda:2025uwo,Alda:2025nsz}, and colliders~\cite{Jaeckel:2012yz,Mimasu:2014nea,
Jaeckel:2015jla,Alves:2016koo,Knapen:2016moh,Brivio:2017ije,Bauer:2017nlg,Mariotti:2017vtv,Bauer:2017ris,
Baldenegro:2018hng,Craig:2018kne,Bauer:2018uxu,Gavela:2019cmq,Haghighat:2020nuh,Wang:2021uyb,deGiorgi:2022oks,
Bonilla:2022pxu,Ghebretinsaea:2022djg,Vileta:2022jou,deGiorgi:2023tvn,Marcos:2024yfm,Arias-Aragon:2024gpm}.

In contrast to the QCD axion, whose mass is generated through non-perturbative effects, the ALP mass is typically treated as a free parameter with $m_a<f_a$ (See Ref.~\cite{Zamoro:2026ily} for a recent investigation of $m_a>f_a$ parameter space). The main goal of this text is to comment of recent results on the ALP mass generation. Indeed, only a few studies exists on this topic, often identifying the ALP with the Majoron~\cite{Chikashige:1980qk,Chikashige:1980ui,Gelmini:1980re} (See Ref.~\cite{Biggio:2026gcs} for a recent review), the Goldstone boson associated with the spontaneous breaking of lepton number (LN).  In these studies, such a mass may arise from the explicit breaking of LN, either via Planck-suppressed operators~\cite{Akhmedov:1992hi,Rothstein:1992rh}, or within specific frameworks addressing the generation of active neutrino masses~\cite{Mohapatra:1982tc,Gu:2010ys,Frigerio:2011in}.

Very recently, Refs.~\cite{deGiorgi:2023tvn,deGiorgi:2024str} have pointed out that a much stronger connections between the ALP mass and other BSM physics is actually possible and in models as simple as the Seesaw mechanisms with additional fermion species, as responsible of the active neutrino masses. In particular, in Ref.~\cite{deGiorgi:2023tvn}, we demonstrated that it is possible of providing the Majoron with a mass in the range $m_a \in [1,\,10^5] \,\keV$, within a specific Type-I Seesaw context~\cite{Minkowski:1977sc,Gell-Mann:1979vob,Yanagida:1979as,Mohapatra:1979ia} with a texture that resemble the low-scale Seesaws~\cite{Kersten:2007vk,Abada:2007ux}. This scenario, however, requires a relatively large decay constant $f_a \in [10^{8},\,10^{12}] \,\GeV$ which follows from the assumption that a single source of LN explicit breaking is responsible simultaneously for generating both the active neutrino masses and the Majoron mass.

\section{The Minimal Massive Majorana Model}\label{MMMModel}

The goal is to connect the Majoron mass with the sources of the active neutrino masses, that by definition represent BSM physics. The idea is to use the LN breaking sources to also break the PQ symmetry, such that the neutrino masses and the Majoron mass are necessarily linked together. In the following, we discuss the embedding of a dynamical SSB mechanism in models with two Heavy Neutral Leptons (HNLs), respecting three ``minimality'' requirements: 
i) only renormalisable interactions are considered in the Lagrangian densities; 
ii) only one large (Majorana) scale is present 
and is associated with the SSB of the PQ symmetry, i.e. only one complex scalar field couples to the HNLs; and 
iii) only one (small) explicit LN and PQ-violating term is introduced. In particular, we assume that there is no other PQ symmetry-breaking term than the one in the Yukawa sector. 
We will see that, once satisfied these three conditions, a unique model that correctly describes the active neutrino 
spectrum and PMNS mixing also predicts a tight correlation between these masses (i.e. associated with the LN breaking) and the Majoron 
mass (i.e. resulting from the PQ breaking).

According to the previous criteria, the SM scalar spectrum is extended only by a single complex scalar field $\phi$, endowed with 
a $U(1)_\text{PQ}$ global symmetry, that gets spontaneously broken by its non-vanishing vev, $f_a$. It is customary to define
\be
\phi\equiv \dfrac{(f_a + \rho)}{\sqrt2} e^{ia/f_a}\,,
\label{phiDEF}
\ee 
being $\rho$ the radial mode and $a$ the GB associated with the SSB of the PQ symmetry, hereafter dubbed as Majoron. The scale 
$f_a$ is assumed to be much larger than the EW scale, $f_a\gg v_\text{EW}$, in such a way that the radial field can be integrated 
out and the Majoron remains the only scalar light degree of freedom at low energies, besides the Higgs. 

Although the RH neutrinos are gauge singlets, they can, in general, transform both under the LN and PQ symmetries and therefore 
can couple both with the SM leptons (and the Higgs) and the scalar field $\phi$. Therefore, when the PQ symmetry gets spontaneously broken the Majorana mass terms for the RH neutrinos are dynamically generated. The manifestation of the LN breaking in the neutral lepton mass matrix can be easily traced by introducing LN spurionic charges. Thus, to identify the specific LN-violating pattern, it is sufficient to impose a specific 
LN charge assignment to the neutral leptons and read the spurionic charges of the terms in the neutral lepton mass matrix. Giving 
a vanishing LN charge to the Higgs simplifies the exercise. 

It is relatively easy to introduce the field $\phi$ in the SS realisations discussed in the previous section, giving rise to 
SSB of the LN. Accordingly to the minimality condition i) of having a renormalisable Lagrangian, we can write
\begin{eqnarray}
-\sL_\text{PQ} = & \phantom{+}\ov{L_L}\,\tH\, Y_N\,  N_R + \ov{L_L}\,\tH\,Y_S\,S_R+ \nn\\ 
& + \frac{1}{2}\,\phi\,\Big[Y_{NN}\,\ov{N_R^c}\,N_R + Y_{SS}\,\ov{S_R^{c}}\,S_R + \label{eq:TypeILNC}\\
&+Y_{NS}\, 
\left(\ov{N_R^{c}}\,S_R + \ov{S_R^{c}}\,N_R\right) \Big] + \hc \,,\nn
\end{eqnarray}
where the Dirac, $Y_{N,S}$, and Majorana, $Y_{NN,NS,SS}$, Yukawa terms are large or small depending on the underlying 
LN symmetry assumed for each scenario. For example, the Linear Seesaw case is obtained, after the EW and PQ SSB, for
\be
Y_S\rightarrow \epsilon Y_{S}\,,\qquad\qquad 
Y_{NN}=0=Y_{SS}
\ee
and identifying
\be
m_{N}= \frac{Y_{N}}{\sqrt{2}} v_\text{EW}\,,\epsilon m_{S}= \frac{\epsilon Y_{S}}{\sqrt{2}} v_\text{EW} \,,\Lambda_{NS} = \frac{Y_{NS}}{\sqrt{2}} f_a \,,
\label{DEFmM1}
\ee
with the ``natural'' hierarchy, from the LN charge assignment point of view, $\epsilon Y_S \ll Y_N$. 

It is straightforward to observe that, besides the LN, the Lagrangian in Eq.~\eqref{eq:TypeILNC} possess an unbroken 
$U(1)_\text{PQ}$ symmetry, with charge assignment
\be
PQ(L_L)=PQ(N_R)=PQ(S_R)=-PQ(\phi)/2\,.
\label{PQChargeAssignmentTypeILNCV}
\ee
As a consequence, the Majoron originated within the PQ SSB remains massless. This can be explicitly seen performing the following field redefinitions:
\be
\chi_L \rightarrow e^{-\frac{i}{2}\frac{a}{f_a}} \chi_L\,,
\label{PQredefinition}
\ee
that remove the Majoron dependence in all the Yukawa terms. Here $\chi_L$ is defined as the vector containing the neutral states,
\be
\chi_L\equiv(\nu_L,\,N_R^c,S^c_R)^T\,,
\label{NeutralVector}
\ee
where $\nu_L \equiv (\nu^e_L, \nu^\mu_L,\nu^\tau_L)$ are the SM neutrinos and $N_R$ and $S_R$ are the two HNLs. The Majoron dependence reappears then in the Lagrangian through the kinetic terms 
as derivative interactions, signalling the underlying presence of the GB shift symmetry.  

One my decide to add explicit PQ breaking terms, that can easily be shown to provide subdominant contributions to the active neutrino masses, but they would be considered as an {\it ad hoc} ingredient to provide the Majoron with a mass --equivalent to an explicit Majoron 
mass term in the scalar potential-- rather than being a common feature of the Majoron and active neutrino mass generation 
mechanisms. For this reason, we do not dub as ``minimal'' this realisation and in particular it violates the minimality 
condition iii) as two independent explicit symmetry-breaking terms are present, one for the LN and the other for the PQ.

We end up identifing the minimal massive Majoron Seesaw (MMM) model where the Majoron mass and the active neutrino masses are indeed correlated. This model satisfies the three minimality conditions i)--iii) and describes realistic active neutrino masses and PMNS mixing. The PQ charges of the fields involved satisfy to 
\be
PQ(L_L) = PQ(N_R) = - \dfrac{PQ(S_R)}{3}=-\dfrac{PQ(\phi)}{2}
\label{PQChargesofMMMmodel}
\ee
and the corresponding PQ conserving and explicitly violating Lagrangian densities read as
\begin{eqnarray}
&&-\sL^\text{MMM}_\text{PQ}=  \ov{L_L}\,\tH\, Y_N\,  N_R +\dfrac{Y_{NN}}{2}\, \phi\,\ov{N_R^c}\,N_R +\label{eq:LagPQ} \\
&&\qquad\qquad\quad
+\frac{Y_{NS}}{2}\,\phi^\ast\, \left(\ov{N_R^{c}}\, S_R + \ov{S_R^{c}}\,N_R\right)+\hc\nn\\
&&-\sL^\text{MMM}_{\epsilon \text{PQ}}= 
\epsilon\ov{L_L}\,\tH\, Y_S\,S_R  + \hc \,,     \label{eq:LagPQV}
\end{eqnarray}
where as usual $Y_N, Y_{NN}$ and $Y_{NS}$ are assumed to be order one, while $\epsilon Y_S$ much smaller. Notice that this Lagrangian is invariant under the interchange of $\phi$ and $\phi^\ast$ as far as the sign of $PQ(\phi)$ in Eq.~\eqref{PQChargesofMMMmodel} is accordingly flipped.

Focussing on the explicit PQ breaking, it is straightforward to check that neglecting the $\epsilon Y_S$ term, the Majoron dependence can be removed from these Yukawa-like interactions, reappearing only in derivative couplings, ending again with a massless Majoron model. Moreover, in this same limit, the active neutrino mass matrix has rank 1 and cannot generate the two observed neutrino mass differences. On the other hand, once this term is taken into consideration, it explicitly breaks both the PQ symmetry and the LN: the Majoron acquires a mass and the active neutrino masses can be described according to the observations, both types of masses being necessarily proportional to $\epsilon Y_S$. 

These equations closely look like the expressions for the Extended Seesaw context~\cite{Lopez-Pavon:2012yda}, and indeed, after the SSB of the EW and PQ symmetries, the resulting lepton mass matrix is given by 
\begin{eqnarray}
&-\sL^\text{MMM}\supset \dfrac{1}{2}\ov{\chi_L}\,\cM_\chi\chi_L^c+\hc\nn\\
&\cM_\chi=\begin{pmatrix}
0 &  m_N & \epsilon\, m_S \\
m_N^T & \Lambda_{NN} & \Lambda_{NS} \\
\epsilon\, m_S^T & \Lambda_{NS} & 0
\end{pmatrix}\,,
\label{MajoronNeutralMassMatrix}
\end{eqnarray}
with the mass terms explicitly given by $\Lambda_{SS}=0$ and
\be
m_{N,S}= \frac{Y_{N,S}}{\sqrt{2}} v_\text{EW}\, , \qquad \Lambda_{NN,NS}= \frac{Y_{NN,NS}}{\sqrt{2}} f_a\,.
\label{DEFmM2}
\ee

We can now use the experimental data from neutrino oscillation experiments, adopting for definiteness the results presented in Ref.~\cite{Esteban:2020cvm} (including the SK atmospheric data) to constrain the parameter space of the MMM model. We scan the parameter space $\Lambda_{NS}$ {\it vs.} $\epsilon Y_S$, running over the different Yukawa couplings taken in the ``natural'' range $|Y_i| \in \left[10^{-2} ,\,  1\right]$, letting $\epsilon$ the only \textit{ad hoc} ``small'' parameter. These conservative conditions show a parameter space where the scale $\Lambda_{NS}$ spans a relatively small range of values, $\Lambda_{NS} \sim 
10^8 - 10^{11}\GeV$, where the Yukawa couplings are larger than $10^{-2}$. As expected, relaxing any of the previous conditions enlarges the parameter space: e.g., requiring the loop contributions to be at most the $30\%$ of the tree-level one, one would allow reaching scales of $\Lambda_{NS} \sim10^{12}\GeV$ 
with Yukawa couplings of order $|Y_i|\sim 0.7$.

\section{The Majoron Mass}\label{MajoronMass}

The simplest approach to calculate the Majoron mass contributions is moving to the chirality preserving basis, leaving the Majoron dependence on the explicit  PQ breaking term, where only the balloon diagram contributes at leading order in the small parameter $\epsilon$.
The leading contribution to the Majoron mass comes from
\begin{eqnarray}
-\sL_{a} \supset&&\dfrac{|m_N|| \epsilon m_S||\eta|}{2\sqrt{M_N\,M_S}(M_N+M_S)}\frac{a^2}{f_a^2} \times\label{FinalChiPreservingBasisRelevantText}\\
&&\quad\times\Bigg(M_N\ov{S_R^c}S_R-M_S\ov{N_R^c}N_R\Bigg)+\hc\,.\nn
\end{eqnarray}
From the couplings in the previous expression, one can calculate the Balloon contribution to the Majoron 
mass in the $\overline{\text{MS}}$ scheme. The result can also be derived employing the CW potential. Assuming $v_{EW} \ll f_a$, in the NO case, the Majoron mass reads
\begin{eqnarray}
m_a^2 =&&\dfrac{ |\eta||m_N|| \epsilon m_S|}{\pi^2 } \dfrac{\sqrt{M_N M_S}}{M_N+M_S}\times\nn\\
&&\times\Bigg[\frac{\left(M_S^2+M^2_N\right)}{f_a^2}\log\left(\dfrac{M_S}{M_N}\right)+ \label{eq:mass-before-constraint}\\
&&\hspace{0.5cm}+ \frac{(M_S^2-M_N^2)}{f_a^2}\left(\log\left(\dfrac{M_N M_S}{\mu^2_R}\right)-1\right)\Bigg]\,. \nn
\end{eqnarray}
Let us notice that in a generic model with explicit PQ symmetry breaking one expects typically $m_a^2 \propto \epsilon f_a^2$.
Instead, with the specific symmetry-breaking pattern introduced for the MMM model, the Majoron mass for large $f_a$ behaves as 
$m_a^2 \propto \epsilon \, v_\text{EW}^2 \log{f_a/\mu_R}$, that is, $m_a^2$ asymptotically depends only logarithmically from the large 
PQ SSB scale, thus allowing a naturally lighter ALP. 

\begin{figure}[t]
\centering
\includegraphics[width=.3\textwidth]{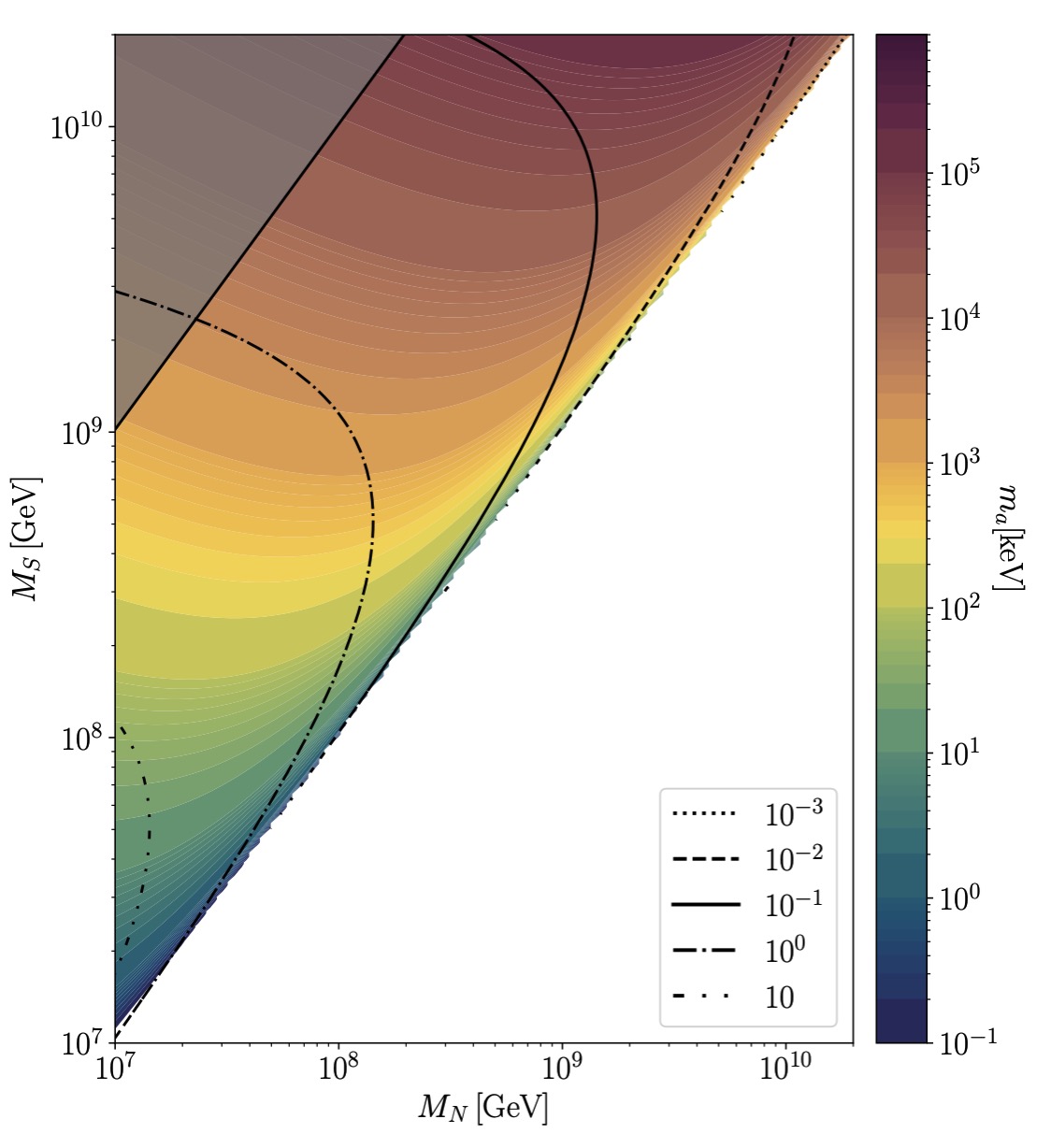}
\caption{\small
\em Dependence of the Majoron mass with respect to the HNL masses $M_{N,S}$, for $f_a=10^{10}\,$GeV and $Y_N=0.01$. The different colours indicate the mass range of the Majoron, while the different black lines are contours of the  $m_L/\sqrt{\Delta m_{\text{atm}}^2}$ ratio. The gray area bounds the region where $m_{T_2}>0.1\,  m_{T_1}$.}
\label{fig:ContourMass}
\end{figure}

To provide an intuitive idea of the Majoron mass behaviour in terms of the relevant model parameters, in Fig.~\ref{fig:ContourMass} 
$m_a$ as a function of the HNLs masses $M_{N,S}$, with the different 
colour nuances indicating the Majoron mass (in keV) as reported on the scale on the right of the figure. The different black 
contours indicate, instead, the corresponding $m_L/\sqrt{\Delta m_{ \text{atm}}^2}$ values as an estimation of the relative size of the 
loop contribution with respect to the tree-level one. This gives an idea of how 
much the parameter space gets constrained for taming the one-loop contribution to the active neutrino masses. In grey, we estimate the region that could induce a $m_{T_2}>0.1 m_{T_1}$. In terms of the 
HNL masses this means that, in order to keep the one-loop vs tree-level ratio below $0.1$, one needs either going close to the mass degeneracy region or increasing $M_S$ above $\sim 10^9$ GeV for a fixed $M_N$, however, this last region would create a larger $m_{T_2}$ contribution. The white region in the left plot of Fig.~\ref{fig:ContourMass} corresponds to the inaccessible region.

\section{Glimpse on the Phenomenology}\label{Pheno}

At tree-level the Majoron 
couples exclusively with neutral leptons, and clearly only the Majoron couplings with active neutrinos can be directly constrained 
by present experiments. Imposing that the observed neutrino oscillation mass differences and PMNS mixing angles are being reproduced by 
opportunely choosing the Dirac and Majorana Yukawas entries of the neutral lepton mass matrix, the lowest order Majoron-active neutrino coupling reads
\be
\sL_{a\nu\nu}=-\dfrac{i\,a}{2f_a}\ov{\nu_L}\, m_\nu \,\gamma_5\,\nu_L^c\,.
\ee 

Couplings of the Majoron with charged SM leptons arise at one loop level through Z and W exchanges. Majoron couplings with quarks are flavour diagonal and therefore relevant bounds from flavour changing neutral observables (for example in $s\to d \,a$ or $b \to d\,a$ transitions) are not expected, 
being suppressed by two loops and by the relatively large scale $f_a$ typically above $10^6$ GeV. Majoron couplings with weak gauge bosons arise at one-loop level but with a $\mathcal{O}(1/f_a^2)$ suppression or at two-loops at $\mathcal{O}(1/f_a)$, therefore they are not phenomenology appealing. Conversely, $\mathcal{O}(1/f_a)$ two-loop contributions 
to the Majoron-photons couplings can be potentially relevant~\cite{Heeck:2019guh}, contributing to the anomalous Lagrangian term. However, the anomalous coupling to photons (as well as to gluons) vanishes in the 
$m_a\to 0$ limit, showing that the lowest order amplitude originates from the $\Box a F\tilde{F}$ effective operator.

\begin{figure}[h!]
\centering
\includegraphics[width=0.4\textwidth]{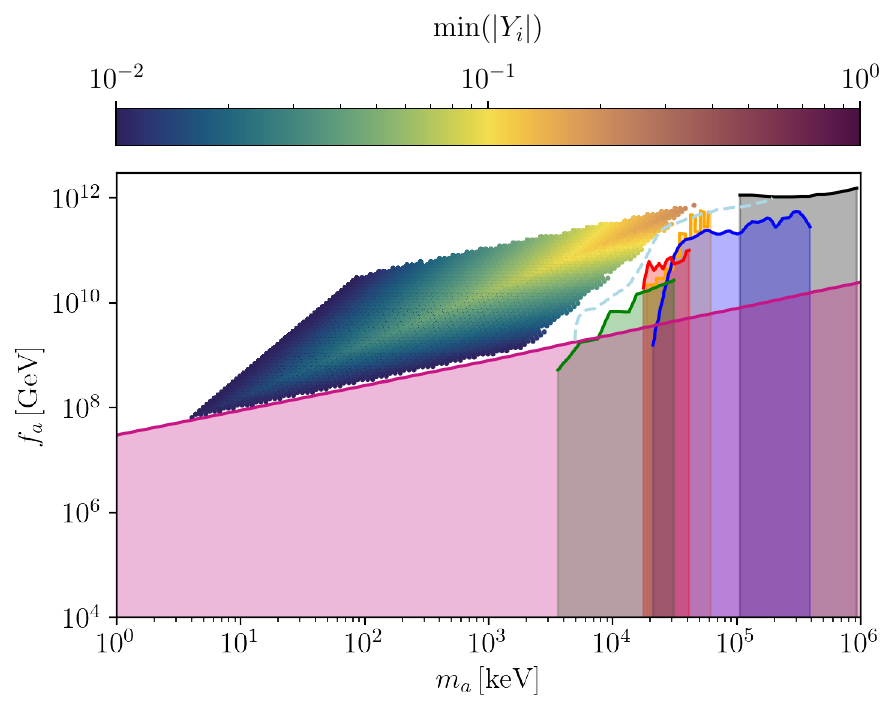}
\caption{\em 
Constraints on Majoron DM. The DM bounds are mainly taken from Ref.~\cite{Akita:2023qiz} and include CMB (\textit{purple area}), 
neutrino experiments, namely Borexino (\textit{green area}), KamLand (\textit{red area}), SK (\textit{blue area}) and projected JUNO (20yr) 
sensitivity (\textit{dashed blue line}). Bounds from Ref.~\cite{Palomares-Ruiz:2007egs}, that include reinterpreted SK data 
(\textit{orange area}) and atmospheric neutrinos data (\textit{gray area}) are also included.}
\label{fig:DM_Majoron}
\end{figure}

The strongest available 
constraints, extracted from Refs.~\cite{Akita:2023qiz, Palomares-Ruiz:2007egs}, are shown in Fig.~\ref{fig:DM_Majoron}, where we refer 
for the detailed labelling. Different colours indicate different $\min (|Y_i|)$ values as shown in the upper bar. As pointed out in 
the literature, from DM-Majoron decays into neutrinos one bounds mainly the SSB scale $f_a$ having 
only a mild dependence on the Yukawa couplings. Notice that by simply requiring the ``naturalness conditions'' (i.e. $\epsilon < 0.01 \times \min(|Y_i|)$ and $|Y_i| \in \left[10^{-2},\,  1\right]$) and taming the loop contribution, our prediction lies just 
above the CMB bounds (purple) and on the left of the current neutrino experiments like SK (blue and orange areas), KamLand (red area) 
and Borexino (green area). Hence, scenarios with ``relaxed naturalness conditions'' are being already ruled out by present neutrino 
data. Future neutrino experiments like JUNO (dashed blue line) could, instead, start probing the region of interest for the MMM model.

\begin{acknowledgments}
I acknowledge partial financial support by the European Union's Horizon 2020 research and innovation programme under the Marie Sk\l odowska-Curie grant agreement No.~101086085-ASYMMETRY and by the Spanish Research Agency (Agencia Estatal de Investigaci\'on) through the grant IFT Centro de Excelencia Severo Ochoa No CEX2020-001007-S and by the grant PID2022-137127NB-I00 funded by MCIN/\hspace{0pt}AEI/\hspace{0pt}10.13039/501100011033.
\end{acknowledgments}


\providecommand{\href}[2]{#2}\begingroup\raggedright
\endgroup
\end{document}